\documentclass[thmsa,a4paper,final]{article}
\usepackage{sw20asas}

\input tcilatex
\QQQ{Language}{
American English
}

\begin{document}

\begin{center}
\textbf{I - THE EXPANDING UNIVERSE}

from

the

\textbf{\ HUGE VOID CENTER:}

\textbf{-}

\textbf{\ THEORY \& MODELLING}
\end{center}

\vspace*{.4in}

- 1st. version: August, 1997 - 2nd. ver.: January, 1998 - 3rd. ver.:
January, 1999 -

by \textbf{Luciano Lorenzi (E-mail: l.lorenzi@senato.it),}

\textbf{former astronomer at the Astronomical Observatory of Turin - Italy -}

\vspace*{.2in}

\begin{center}
$\mathbf{ABSTRACT}$
\end{center}

\begin{quote}
\textbf{To add to and refine the model presented at the Grado3 meeting, of a
radial Hubble expansion from the Bahcall \& Soneira huge void center, it is
here outlined a formal analytical formulation of the theory and modelling on
which the previous check work based itself. The new Hubble Law is }:$\mathbf{%
\ }$%
\[
\mathbf{\dot r=Hr+\Delta H\cdot (r-R}\cos \mathbf{\gamma )+R\dot w}\sin 
\mathbf{\gamma }
\]
\textbf{The meaning and expression of the total }$\Delta H$ \textbf{have
been obtained and examined through a Galaxy Hubble law analysis based on
derivatives with respect to light-space. This }$H^{\prime }s$ \textbf{%
variation, as predicted by the model, is due to a combination of two
dominant effects, respectively time (TE) and space (SE), . Structurally the
scattering noise of the nearby Universe Hubble ratios seems to be partially
caused by the perturbative term }$R\dot w\sin \gamma /r$\textbf{. In
conclusion a fundamental confirmation test of the model is presented; as a
physical result one supports the density formula:}

\[
\mathbf{\rho }_0\mathbf{>3H}_0^2\mathbf{/2\pi G} 
\]

\newpage\ 
\end{quote}

\section{INTRODUCTION}

It's well known that so far a lot of time has been devoted to the problem of
the Hubble constant; indeed this has become one of the most controversial
and crucial scientific questions in the astronomical debate of the century.
But in the last decades the investigations about value and meaning of $H$
have greatly increased within different schools of thought, in particular
after the enlightening observation of Rubin, Ford and Rubin (1973), the
so-called RFR effect. This was the starting point of a long study, that
through the fundamental contribution of a lot of works in observational
cosmology, led the author (Lorenzi, 1989-91-93-94-95) to acknowledge an
expansion center in the huge void of Bahcall \& Soneira (1982) (Lipovetsky,
1987) who, following up the pioneering detection by Kirshner et al. (1981)
of an apparent absence of relatively bright galaxies ($z\approx 0.04-0.06$)
over an enormous volume in Bootes, discovered and described a $\sim 300$ $%
Mpc $ void of catalogued nearby rich Abell clusters of galaxies in the
direction $l^{II}\approx 140^0-240^0,b^{II}\approx 30^0-50^0$, with $%
z\approx 0.03-0.08 $ . Here theory and modelling of all the research has
been re-run, beginning from the original toy-model, proceeding through a few
fundamental mathematical developments, and concluding with remarkable
consequences for the Hubble ratio behaviour and the cosmography and
evolution of the Universe.

\newpage\ 

\section{THE ORIGINAL TOY-MODEL}

\subsection{Preliminary remarks}

The father of the Big Bang, the Belgian abbot Georges Lemaitre, originally
imagined a different situation to the one worked out by Gamow (1948),
Alpher, Herman (1948). He hypothesized the primordial explosion as being
caused by a sort of fission cosmo-bomb, whose fragments started the journey
producing the present dilution of the Universe.

Matching the idea of Lemaitre (1933a,b) with the conception of Ambartsumian
(1961), by which the protogalaxies (and all the observed matter) could have
come out of extremely small concentrated bodies, it's possible to give one
intuitive explanation for the appearances of the most remote objects of the
Universe, the Quasars, which have the peculiarity, as well known, of
appearing extremely bright and small in the sky. Anyway, if we think
plausible that a ''Hot Hard Big Bang'' may have occurred, following some
elementary spherical symmetry, the observation of the Universe from any
splinter might show a few peculiarities, and possibly allow to locate a
preferred relative point, that is the ''expansion center''.

Let us remember the contributions of Hubble (1929) and Penzias \& Wilson
(1965), who gave crucial observation evidence for the Big Bang; but also the
question of isotropy or anisotropy, still rather dimmed by the authoritative
Cosmological Principle (Milne, 1933). A very stimulating astronomical
observation, though it was not equally meaningful in statistical terms
(Sandage \& Tamman, VI-1975b), was surely the one pointed out by Rubin, Ford
and Rubin (1973); it raised many perplexities about the full validity of
Hubble law (Nottale,Pecker,Vigier,Yourgrau, 1976). From that time the RFR
effect has become the object of many important surveys, which have closely
examined and partially interpreted a few observed anisotropies (see ref. in
Lynden-Bell et al., 1988).

\subsection{Introduction to an elementary Big Bang mechanics}

In the full awareness that the greatest singularity of all, that referring
to time ''0'', cannot be analysed otherwise than by conjectures, a simple
toy-model of Big Bang mechanics was developed by the author (1989), the
starting point of which is the Lemaitre fragmentation hypothesis, out of the
quantic and relativistic extremes. Such approach bases itself on the
fundamental principle of dynamics. Let us imagine a spherical cosmo-bomb, a
relative explosion according to the $spherical$ $symmetry$ and the derived
impelling bomb force as the only one acting force simply proportional to the
mass of the inner globe. In other words we might hypothesize a coefficient $%
\varepsilon $ which should just represent the impelling force developed by
the unitary mass and representing the exclusive action force at the
beginning. In consequence of a simultaneous explosion in all the globe
points, assuming also the instantaneous propagation of the involved shock
wave, we can reasonably consider as maximum the impulse received by the
first shell on the outside surface. The impulses taken up by each internal
shell, submitted to the burst impact of a minor globe, would become smaller
and smaller. Therefore the impulse distribution should go from the zero
value in the globe centre to a maximum value on the surface, while the
impelling force acting on different shells should always be proportional to
the mass of the involved inner globe. Now let us consider one spherical
shell, of radius $r_b$ , and write Newton law as 
\begin{equation}
F=\frac{dp}{dt}  \label{1}
\end{equation}
where the impelling force $F$ for the chosen shell is assumed constant as 
\begin{equation}
F=\frac 43\pi r_b^3\rho \varepsilon  \label{2}
\end{equation}
Of course, in eq. (2) $\rho $ would represent the cosmic matter density in
its primordial state, $\varepsilon $ the above mentioned speculative
coefficient. Eq. (1), with $p=mv$ , becomes 
\begin{equation}
m\Delta v=F\Delta t  \label{3}
\end{equation}
$\Delta t$ is the action time of $F$ ; $\Delta v=v_0$ the acquired expansion
velocity of the shell as a whole during the explosion time $\Delta t$ in
which, it is important to underline that, there is no other acting force
outside the speculative $F$ of eq. (2); $m$ the mass of the shell itself $%
\Delta r_b$ thick, that is 
\begin{equation}
m=4\pi r_b^2\Delta r_b\rho  \label{4}
\end{equation}
So eq. (3) gives automatically 
\begin{equation}
v_0=Cr_b\hspace{.5in}with\hspace{.5in}C=\frac{\varepsilon \Delta t}{3\Delta
r_b}\rightarrow \infty  \label{5}
\end{equation}
The previous equation refers to the starting radial velocity $v_0$ of any
splinter with respect to the primordial globe centre, in a simple Big Bang
hypothesis of a primordial shattering (according to the spherical symmetry)
produced by an impelling force proportional to the mass of the inner globe,
and assuming also the identity $"splinters=protogalaxies".$ Such velocity $%
v_0$ is proportional to the distance $r_b$ from the centre and to a function 
$C$ of constant value and $\rightarrow \infty $ at least in the same way of
the impelling coefficient $\varepsilon $.

Now, classical kinematics furnishes radial run and velocity, $R$ and $\dot R$
, of any splinter after a proper period $"t"$ from the end of the Big Bang $%
t=0$. In fact, immediately after the exclusive explosive action, it is
possible to fix at $t=0$ the beginning of the normal gravity deceleration.
So, considering such deceleration $\ddot R$ as $d\dot R=\ddot Rdt$, it
follows from the mean theorem: $\int_{v_0}^{\dot R(t)}d\dot R=\int_0^t\ddot
Rdt=\langle \ddot R\rangle t$, that is $\dot R(t)=v_0+\langle \ddot R\rangle
t$ , whose further integration gives the radial run $R$ as : 
\begin{equation}
R=r_b+v_0t+\frac 12\langle \ddot R\rangle t^2  \label{6}
\end{equation}
and then we can write in sequence

\[
R=r_b(1+Ct)+\frac 12\langle \ddot R\rangle t^2 
\]
\[
r_b=\frac 1{1+Ct}(R-\frac 12\langle \ddot R\rangle t^2) 
\]
\[
\dot R=Cr_b+\langle \ddot R\rangle t=\frac C{1+Ct}R-\frac C{1+Ct}\frac
12\langle \ddot R\rangle t^2+\langle \ddot R\rangle t 
\]
\[
\lim_{C\rightarrow \infty }\frac C{1+Ct}=\frac 1t 
\]
\begin{equation}
\dot R=\frac Rt+\frac 12\langle \ddot R\rangle t  \label{7}
\end{equation}
Above we have the splinter radial velocity as function of $t,R$ and of the
medium radial deceleration, $\langle \ddot R\rangle $, to which we need to
apply a correct dynamic formula. According to classical mechanics for an
orbiting splinter-galaxy of negligible mass $m$ is 
\begin{equation}
\ddot R-R\dot \vartheta ^2=-\frac{GM}{R^2}  \label{8}
\end{equation}
where $R\dot \vartheta ^2=a_c$ is representing the hypothetical centripetal
acceleration due to an undefined angular velocity $\dot \vartheta $, that
could be roughly considered as belonging to a same thick orbital plane only
for the nearby Galaxy environment, we can rewrite eq. ( 8) according to the
Newtonian formulation of the first Einstein equation as below: 
\begin{equation}
\ddot R=-\frac 43\pi G\rho (t,R)R(t)+\dot \vartheta ^2(t,R,...)R(t)
\label{9}
\end{equation}
At present it is not easy to imagine the physical meaning of such $a_c$ ;
however it does not affect the procedure to obtain our aim, and its
inclusion has to have only an explanatory and qualitative meaning.
Therefore, after integrating eq. (9) according to the mean theorem, we can
obtain the following 
\begin{equation}
\langle \ddot R\rangle =-\frac 43\pi GR_0\rho _{*}  \label{10}
\end{equation}
having fixed mathematically 
\begin{equation}
\rho _{*}=\left[ \langle \rho \rangle -\langle \dot \vartheta ^2\rangle
\frac 3{4\pi G}\right] \frac{\int_0^{t_0}R(t)dt}{R_0\int_0^{t_0}dt}
\label{11}
\end{equation}
and being at the same time 
\begin{equation}
\rho _{*}\leq \frac{\langle \rho \rangle \int_0^{t_0}R(t)dt}{%
R_0\int_0^{t_0}dt}  \label{12}
\end{equation}
where, of course, the sign $=$ implies $\langle \dot \vartheta ^2\rangle =0$.

The (10) formula represents an acceptable expression for the fixed average
radial deceleration, $\langle \ddot R\rangle $ , of all the
splinter-galaxies which should be at the same distance $R_0$ , at our epoch $%
t_0$, from the centre of the Big Bang sphere;\textbf{\ }$\rho _{*}$\textbf{\
is a density\TEXTsymbol{\backslash}rotation function} which referees to the
average density $\langle \rho \rangle $ of the sphere expanded to $R_0$ at $%
t_0$ and to the proper average square hypothetical rotational velocity $%
\langle \dot \vartheta ^2\rangle $ , hence we can reasonably write $\rho
_{*}=\rho _{*}(t_0,R_0,...)$; $G$ is the gravitation constant. Substituting
(10) in (7), we finally obtain 
\begin{equation}
\dot R=R_0\left[ \frac 1{t_0}-\frac 23\pi Gt_0\rho _{*}(t_0,R_0)\right]
\label{13}
\end{equation}
that is to say 
\begin{equation}
\dot R=H_{0_{s^{-1}}}R_0  \label{14}
\end{equation}

Such $\dot R=H_{s^{-1}}R$ is a true\textbf{\ radial Hubble law}, in $c.g.s$.
units, \textbf{centred on the expansion center}, with the following Hubble
constant formulation: 
\begin{equation}
H_{s^{-1}}=\frac 1t-\frac 23\pi Gt\rho _{*}(t,R)  \label{15}
\end{equation}

\section{ GALAXY HUBBLE CONSTANT VARIATION: $\Delta H_{MW}$}

The Hubble law in (14) can be applied to our Galaxy, of course, being $R_0$
its distance from the origin at the epoch $t_0$, $\dot R$ the involved
recession velocity, and $H_{0_{s^{-1}}}=H_{0_{s^{-1}}}(t_0,\rho
_{*}(t_0,R_0(t_0)))=H_{0_{s^{-1}}}(t_0,R_0(t_0))$ a function which can be
considered constant in our epoch $t_0$, all over the sphere having radius $%
R_0$.

The total derivative with respect to time of $H_{s^{-1}}$,whose units are $%
s^{-1}$, is then 
\begin{equation}
\frac{dH_{s^{-1}}(t_0,\rho _{*})}{dt_0}=\frac{\partial H_{s^{-1}}}{\partial
t_0}+\frac{\partial H_{s^{-1}}}{\partial \rho _{*}}\frac{d\rho _{*}}{dt_0}
\label{16}
\end{equation}
being by (15) 
\begin{equation}
\frac{\partial H_{s^{-1}}}{\partial t_0}=-\left\{ \frac 2{t_0^2}-\frac{%
H_{0_{s^{-1}}}}{t_0}\right\} \hspace{1.0in}\frac{\partial H_{s^{-1}}}{%
\partial \rho _{*}}=-\frac 23\pi Gt_0  \label{17}
\end{equation}
So, applying the Taylor series, it results 
\begin{equation}
\Delta H_{s^{-1}}=-\left\{ \frac 2{t_0^2}-\frac{H_{0_{s^{-1}}}}{t_0}+\frac
23\pi Gt_0\frac{d\rho _{*}}{dt_0}\right\} \Delta t_0+...  \label{18}
\end{equation}

Now we proceed to transform the previous relations in Hubble units. For this
purpose let us begin to consider $\Delta t_0$ as the number of light-seconds
of the distance $r$. In other words any luminous signal, owing to the finite
speed of light $c$, reaching us at the epoch $t_0^{^{\prime }}$ with a delay
with respect to the emission epoch $t_0^{^{\prime \prime }}<t_0^{^{\prime }}$%
, will have covered during that time the distance $r=-c(t_0^{^{\prime \prime
}}-t_0^{^{\prime }})$, that is 
\begin{equation}
r=\frac{\delta r}{\delta t_0}\Delta t_0=-c\Delta t_0  \label{19}
\end{equation}

\textbf{In (19) we have used }$\delta r$\textbf{\ to indicate the
infinitesimal space run by light travelling towards us during an
infinitesimal }$dt_0=$\textbf{\ }$\delta t_0$ \textbf{of our past time. Such
indication is important only to avoid confusion with the conventional }$dr$%
\textbf{, which being included in }$\dot r$\textbf{\ represents the
infinitesimal distance variation of any galaxy observed at the light
distance }$r$.\textbf{\ Consequently we define here, in place of the usual
total derivative with respect to time, an alternative total derivative, as }$%
\delta /\delta r$, \textbf{computed with respect to light-space.}

Now, indicating $r$ in $Mpc$ and writing $H$ in $Km$ $s^{-1}Mpc^{-1}$ it is: 
\begin{equation}
\Delta t_0=\frac{r_{cm}}{-c}=-1.029\times 10^{14}\cdot r_{Mpc}\hspace{1.0in}%
H_{s^{-1}}=3.24\times 10^{-20}H  \label{20}
\end{equation}

The incremental variation (18), after the above transformation, becomes 
\begin{equation}
\Delta H_{MW}=3.17\times 10^{33}\left[ \frac 2{t_0^2}-\frac{H_{0_{s^{-1}}}}{%
t_0}+\frac 23\pi Gt_0\frac{d\rho _{*}}{dt_0}\right] \cdot r+.....  \label{21}
\end{equation}

Here the total derivative of the composed function $\rho _{*}$ with respect
to time concurs to define the Hubble constant variation $\Delta H_{MW}$ of
our Galaxy, being in particular $\frac{d\rho _{*}}{dt_0}<0$ if the Universe
is expanding with $d\rho _{*}<0$ and $dt_0>0$ . Consequently, after having
assumed 
\begin{equation}
K_0=\left( \frac{\delta H_{MW}}{\delta r}\right) _{r=0}=3.17\times
10^{33}\left[ \frac 2{t_0^2}-\frac{H_{0_{s^{-1}}}}{t_0}+\frac 23\pi Gt_0%
\frac{d\rho _{*}}{dt_0}\right]  \label{22}
\end{equation}
from (21) one obtains 
\begin{equation}
H_{MW}(r)\cong H_0+K_0r+...  \label{23}
\end{equation}
that is the Milky Way Hubble constant trend as function of the light-space $%
r $ in megaparsecs.

\section{THE GALAXY HUBBLE LAW: $\dot R_{MW}=H_{MW}R_{MW}$}

Here we are able to show the complete mathematically auto-consistency of the
model of a radial Hubble expansion, based on the fundamental equation (24) 
\begin{equation}
\left( \frac{dR_{\left( Km\right) }}{dt}\right) _{MW}=H_{MW}R_{MW}
\label{24}
\end{equation}
that is on a pure Hubble law which, instead of being centred onto the
Galaxy, has been here shifted into the center of the huge void of Bahcall \&
Soneira (Lorenzi, 1991-1996), at the distance $R_{MW}$ from the Milky Way $%
(MW)$. So eq. (24) represents the radial velocity of our Galaxy, whose well
known Hubble constant at our epoch $t_0$ takes the value $H_{MW}=H_0$. In
the previous equation we can substitute $dt=dt_0$ with $dt_0=-\delta r/c$
(see eq. 19), $dt_0$ being the negative number of light-seconds
corresponding to the distance $\delta r$ covered in the past by the light
emitted by any hypothetical observed source, and $c$ the speed of light,
travelling towards the earth, in $Km$ $s^{-1}$ units of course. Immediately
it follows 
\begin{equation}
\frac{\delta R_{MW}}{\delta r}=-\frac{H_{MW}R_{MW}}c  \label{25}
\end{equation}
that, in first order of approximation for $r\rightarrow 0$, can be solved as 
\begin{equation}
R_{MW}(r)\cong R_0+q_0r+...  \label{26}
\end{equation}
where 
\begin{equation}
q_0=-\frac{H_0R_0}c  \label{27}
\end{equation}

Let us note the dimensionless number $q_0$, and consequently the possibility
to adopt directly the $Mpc$ units for $R,r,$ as it is for the ratio $c/H_0$.
Practically eq. (26), in which both $R_0$ and $q_0$ are here constant
quantities, gives us the linear trend of $R_{MW}$ versus $r$, that is the
distance $R$, of our galaxy from the assumed expansion center, corresponding
to any covered light-distance $r$, therefore corresponding to the epoch of
the light emission by the observed source; hence eq. (26) gives us the
variation with time of the Milky Way radial distance $R_{MW}$ from the void
center. Of course $R_0$ is our Galaxy $R_{MW}$ at our epoch $t_0$. Deriving
(25) again, we have 
\begin{equation}
\frac{\delta ^2R_{MW}}{\delta r^2}=-\frac{R_{MW}}c\frac{\delta H_{MW}}{%
\delta r}-\frac{H_{MW}}c\frac{\delta R_{MW}}{\delta r}  \label{28}
\end{equation}
representing here a different way to indicate the deceleration $\ddot R$. So
it follows: 
\begin{equation}
\left( \frac{\delta H_{MW}}{\delta r}\right) _{r=0}=\frac{H_0^2}c-\frac
c{R_0}\left( \frac{\delta ^2R_{MW}}{\delta r^2}\right) _{r=0}=K_0  \label{29}
\end{equation}
from which, still in first order of approximation for $r\rightarrow 0$, we
derive again the (23) equation.

\subsection{BY A SIMULATION}

At this point we have two theoretical relations, (23) and (26), that
practically represent the values of $H_{MW}$ and $R_{MW}$ as functions of
time. Consequently $\dot R_{MW}=H_{MW}R_{MW}$ can now be integrated within
the limits of a simulation carried out by adopting as rigorously true the
previous equations (23) and (26) (these being so for $r\rightarrow 0$), as
was previously done in the contribution presented at the Sesto Pusteria
International Workshop (Lorenzi, 1995b,c). Therefore :

\begin{equation}
\int_0^{R_0}dR_{\left( Km\right) }=\int_0^{t_0}H_{MW}R_{MW}\cdot dt_0
\label{30}
\end{equation}

being

\begin{equation}
R_{MW}=R_0\rightarrow r=0\rightarrow t_{R=R_0}=t_0\text{ }(our\text{ }epoch)
\label{31}
\end{equation}
\begin{equation}
R_{MW}=0\rightarrow r=-\frac{R_0}{q_0}\rightarrow t_{R=0}=0\text{ }(adopted%
\text{ }zero\text{ }time)  \label{32}
\end{equation}

So one obtains:

\begin{equation}
cR_0=-\int_{-\frac{R_0}{q_0}}^0(H_0+K_0\cdot r)(R_0+q_0\cdot r)\cdot \delta r
\label{33}
\end{equation}
where, having imposed $q_0=-\frac{H_0R_0}c$, the $K_0$ of (29) assumes an
appropriate simulation value. It follows 
\begin{equation}
c=\int_0^{\frac c{H_0}}(H_0+K_0\cdot r)(1-\frac{H_0}cr)\cdot \delta r
\label{34}
\end{equation}
whose solution, within the limits of the previous simulation, results to be
finally 
\begin{equation}
\mathbf{K}_0\mathbf{=}\left( \frac{\mathbf{3H}^2}{\mathbf{c}}\right) _{%
\mathbf{r=0}}\mathbf{=}\left( \frac{\mathbf{\delta H}}{\mathbf{\delta r}}%
\right) _{\mathbf{r=0}}  \label{35}
\end{equation}

What expressed in Eq. (35) has indeed general validity in time; so it is
possible correctly to carry out the integration

\begin{equation}
\int_{H_0}^{H_{MW}}\frac{\delta H}{H^2}=\frac 3c\int_0^r\delta r  \label{36}
\end{equation}
whose solution gives:

\begin{equation}
H_{MW}=H_0+\frac{3H_0^2}{c-3H_0r}r  \label{37}
\end{equation}
that is the following formulas to $H_{MW}$ and $K_{MW}$

\begin{equation}
H_{MW}=\frac{H_0c}{c-3H_0r}\hspace{.4in}\Delta H_{MW}=\frac{3H_0^2r}{c-3H_0r}%
\hspace{.4in}K_{MW}=\frac{3H_0^2}{c-3H_0r}  \label{38}
\end{equation}

Analogously the eq. (25), where now the $H_{MW}$ function is known, can be
integrated to find the $R_{MW}$ formula. The solution of the integral

\begin{equation}
\int_{R_0}^{R_{MW}}\frac{\delta R_{MW}}{R_{MW}}=-\int_0^r\frac{H_0}{c-3H_0r}%
\delta r  \label{39}
\end{equation}
after the logarithmic reduction, gives finally:

\begin{equation}
R_{MW}=R_0\left( 1-\frac{3H_0r}c\right) ^{\frac 13}  \label{40}
\end{equation}

The last considerable result to be drawn is now that referring to the Galaxy 
\textbf{radial deceleration} formula (28), which at our epoch $t_0$, after
the introduction of the appropriate derivatives, becomes:

\begin{equation}
\left( \frac{\mathbf{\delta }^2\mathbf{R}_{MW}}{\mathbf{\delta r}^2}\right)
_{\mathbf{r=0}}\mathbf{=-2}\frac{\mathbf{H}_0^2\mathbf{R}_0}{\mathbf{c}^2}
\label{41}
\end{equation}
or, in $c.g.s.$ units, the equivalent one:

\begin{equation}
\ddot R_{MW_{t=t_0}}=\left( \frac{d^2R_{cm}}{dt^2}\right) _{MW_{t=t_0}}=-2%
\text{ }H_{s^{-1}}^2(t_0)\cdot R_{0_{cm}}  \label{42}
\end{equation}

Finally, by substituting the previous deceleration expression in eq. (9), as
a physical result one obtains that at our epoch the inner Universe including
the whole huge void of Bahcall \& Soneira has a matter density $\rho _0$ ,
with a lower limit, according to the following simple formula: 
\begin{equation}
\mathbf{\rho }_0>\frac{\mathbf{3H}_{s^{-1}}^2\mathbf{(t}_0\mathbf{,R}_0%
\mathbf{)}}{\mathbf{2\pi G}}  \tag{42b}
\end{equation}

\section{THE EXPERIMENTAL MODEL}

In order to realize the operative purposes of the model, letting a side for
the moment all the above developments, one may rather consider the only
fundamental Eq. (14) as the general Hubble law 
\begin{equation}
\dot R=HR  \label{43}
\end{equation}
which can be applied to any place in the Universe, according to the
Cosmological Principle, but taking into account the possibility that the
assumed homogeneous and isotropic expansion is perturbed by local effects.
In particular the choice could be that previously adopted (Lorenzi, 1991) of
applying Eq. (43) to the void center (VC) of the huge void of Bahcall \&
Soneira (1982), because the void itself is a deviation from the local
homogeneity and isotropy, that is from the environmental standard
conditions. This void (hereafter BSHV), extending $100^{0\text{ }}$across
the sky in the redshift range of $z\approx 0.03-0.08$, is centered
approximately at $\alpha _{VC}\approx 9^h,\delta _{VC}\approx
+30^0(l_{VC}\approx 195^0,b_{VC}\approx +40^0)$, and appears to extend, in
projection, $300$ $h^{-1}Mpc$ by $\geq $ $60$ $h^{-1}Mpc$ (Bahcall, 1988);
consequently such a highly under-dense region dominates our sky. But there
are other important scientific references, which seem to highlight the
cosmological and cosmographic meaning of the BSHV and its center, VC. In
particular we refer to the VC position belonging to the hemisphere with
smaller Hubble ratios, inside Region 1 of Rubin, Ford and Rubin (RFR effect:
1973); to the detected Optical Dipole of Lahav (1987)(LOD: $l=227^0\pm
23^0,b=+42^0\pm $ $8^0$), which follows from about 15000 optical galaxies at
the low average depth of $50h^{-1}Mpc$ practically in the same direction of
VC; to the detection by Geller \& Huchra (1989) of the ''Great Wall''
surrounding the BSHV with a minimum extent of $60h^{-1}Mpc\times
170h^{-1}Mpc $; and, finally, to the observed MBR dipole (COBE: Smoot et
al., 1992).

That being stated, the basic hypothesis of research became, of course, that
of a radial expansion, whose formulation (43) follows a pure Hubble law
shifted into the center of the void (VC). The experimental model, here
re-examined, is exclusively geometric, of the Euclidean type, within the
limits of the present non relativistic observational cosmology $(cz\ll c)$.

\subsection{Analytical solution}

Let us consider in Eq. (43) $R$ as representing the distance of a generic
galaxy/group/cluster from VC, $\dot R$ the involved radial velocity, $H$ the
corresponding Hubble flow parameter; and Let us describe a generic
perturbation by means of the following mathematical differential: 
\begin{equation}
d\dot R=H\cdot dR+R\cdot dH+\delta  \label{44}
\end{equation}

Let us jointly consider the trigonometrical distance $r$ 
\begin{equation}
r^2=R^2+(R+dR)^2-2R(R+dR)\cos w  \label{45}
\end{equation}

between any two galaxies respectively $R$ and $R+dR$ distant from VC (see
Fig. 1 of the vectorial solution). Such value of $r$ in (45) represents the
distance covered from a galaxy to another by the luminous signal that,
according to special relativity, travels all the time with a constant speed
of light with respect to the receiver galaxy. Consequently the source
distance $r$, registered by the observer and computed in \textbf{light-space}%
, has to be considered to be the same as that of the source at the epoch of
light emission, in which the galaxies find themselves at the distances $R$
and $R+dR$ from the void center VC, respectively.

If now we insert (43) and (44) into the derivative of (45) (as in the
Appendix of the 1991 paper), with the second order differential $\delta $
fixed $=0$ as first step of Eq. (44) applied to the very nearby Universe,
and assume the basic hypothesis $\dot w=0$ concerning the angle $w$ between
two radial runs of the radial expansion from VC, that is with exclusion of
any differential rotation, these mathematical steps follow:

\[
2r\dot r=4R\dot R+2dRd\dot R+2\dot RdR+2Rd\dot R-4R\dot R\cos w-2\dot
RdR\cos w-2Rd\dot R\cos w 
\]
\begin{eqnarray*}
r\dot r &=&2HR^2+HdR^2+RdHdR+HRdR+HRdR+dHR^2+ \\
&&\ -2HR^2\cos w-HRdR\cos w-HRdR\cos w-dHR^2\cos w
\end{eqnarray*}

\[
r\dot r=H(2R^2+dR^2+2RdR-2R^2\cos w-2RdR\cos w)+R^2dH-R^2dH\cos w+RdHdR 
\]

\[
r\dot r=Hr^2+RdHdR+R^2dH(1-\cos w) 
\]

\[
\dot r=Hr+RdH\left[ \frac{dR+R(1-\cos w)}r\right] 
\]

\[
R=(R+dR)\cos w+r\cos \gamma 
\]

\[
\cos w=\frac{R-r\cos \gamma }{R+dR} 
\]

\[
r^2=R^2+(R+dR)^2-2R(R-r\cos \gamma )=dR^2+2RdR+2Rr\cos \gamma 
\]

\[
\frac{dR+R(1-\cos w)}r=\frac{dR^2+2RdR+rR\cos \gamma }{r(R+dR)}=\frac{%
r^2-rR\cos \gamma }{r(R+dR)} 
\]

\begin{equation}
\dot r=Hr+RdH\left[ \frac{r-R\cos \gamma }{R+dR}\right]  \label{46}
\end{equation}

or 
\begin{equation}
\frac{\dot r}r=H+R\frac{dH}{dR}X=H+R\frac{dH}rY  \label{47}
\end{equation}
with 
\begin{equation}
X=\frac{1-(R/r)\cos \gamma }{1+R/dR}\hspace{.3in}Y=\frac{r-R\cos \gamma }{%
R+dR}\hspace{.3in}dR=-R+\sqrt{R^2+r^2-2rR\cos \gamma }  \label{48}
\end{equation}

Eq. (46), when applied to human-made measurements, represents a modified
formula for the local Hubble law. It holds when it is $\dot w=0$ and $\delta
=0$. Otherwise, taking into account $\dot w\neq 0$ , the further processing
of (46) and the application of the sine theorem complete the previous
formulation of radial velocity, through the addition of a new term. Finally
it results: 
\begin{equation}
\mathbf{\dot r=Hr+dH\cdot (r-R}\cos \mathbf{\gamma )+R\dot w}\sin \mathbf{%
\gamma }  \label{49}
\end{equation}

being by definition $dR$ the analytical differential, and so 
\begin{equation}
\lim_{dR\rightarrow 0}\left[ \frac{r-R\cos \gamma }{1+dR/R}\right] =r-R\cos
\gamma  \label{50}
\end{equation}

Now we have to extend all the above differential process to the more
realistic case of $\delta \neq 0$ in (44).

Indeed if $\Delta H$ and $\Delta R$ are considered as finite differences,
rather than as the differentials $dH$ and $dR$, the (44) can be rewritten as
follows 
\begin{equation}
\Delta \dot R=(H+\Delta H)(R+\Delta R)-HR=H\Delta R+R\Delta H+\Delta H\Delta
R  \label{51}
\end{equation}

Substituting such finite difference $\Delta \dot R$ to $d\dot R$ in the
previous development (46), also including $\dot w\neq 0$, we'll finally find
the following finite difference equation 
\begin{equation}
\mathbf{\dot r=Hr+\Delta H\cdot }\left[ \frac{r-R\cos \gamma }{R+\Delta R}%
\right] \mathbf{\cdot (R+\Delta R)+R\dot w}\sin \mathbf{\gamma }  \label{52}
\end{equation}
that is 
\begin{equation}
\mathbf{\dot r=Hr+\Delta H\cdot (r-R}\cos \mathbf{\gamma )+R\dot w}\sin 
\mathbf{\gamma }  \label{53}
\end{equation}

Eq. (53), with the cancellation of $R+\Delta R$ in (52), actually shows the
possibility of existence of large $\Delta R$ in the model characterized by
small finite $\Delta H^{\prime }s$; so it coincides formally with that
obtained in (49) by adopting $dR\rightarrow 0$ as a consequence of $\delta
=0 $.

In conclusion Eq. (53) results to be our searched new Hubble law. In it $r$ $%
=r_{ga}$ is the classical separation, in terms of light-space observed at
our epoch and referring to the epoch of light emission, between the Milky
Way ($MW$) and other galaxies $(ga)$ ; $\dot r=\frac{dr}{dt}$ is its
variation in time as registered by us in our epoch and connected to the
appropriate $\Delta H$ ; $\gamma $ is the observed angle at $MW$ between the
direction of the reference point VC and the direction referring to the past
galaxy/group/cluster position observed. $\gamma $ , being here $\alpha
,\delta $ known equatorial coordinates, can be calculated as follows: 
\begin{equation}
\cos \gamma =\sin \delta _{VC}\sin \delta +\cos \delta _{VC}\cos \delta \cos
(\alpha -\alpha _{VC})\hspace{.7in}\sin \gamma =(1-\cos ^2\gamma )^{1/2}
\label{54}
\end{equation}
Finally we shall note how $\Delta H=0$ and $\dot w=0$ reduce eq. (53) to the
canonical Hubble law $\dot r=Hr$.

\subsection{Vectorial verification}

The previous analytical solution is clearly consistent with a finite
difference scenario like that graphically represented in Figure 1, where the
small increment $\Delta H$ has to be referred to a galaxy/group/cluster
that, at the epoch of emitted light we now receive, is respectively $r$
distant from the Milky Way and $R+\Delta R$ far from the expansion center VC.

Let us try a vectorial approach in order to verify the obtained solution
more intuitively. In this case one can consider the radial velocity $\dot r$
of the emission epoch as the difference of expansion velocity projected on
the radial direction $r$ , between any galaxy/group/cluster and our Milky
Way. So, starting as always from the basic hypothesis of radial expansion $%
\dot R=HR$ with the assumption $\dot w=0$, it is easy to write

\begin{equation}
\dot r=(H+\Delta H)(R+\Delta R)\cos \alpha -HR(-\cos \gamma )  \label{55}
\end{equation}
where the angle $\alpha $ between $r$ and $R+\Delta R$ follows the simple
equality

\begin{equation}
\alpha =180-w-\gamma  \label{56}
\end{equation}

In (55) $(R+\Delta R)\cos \alpha $ can be transformed trigonometrically as
follows:

\begin{equation}
(R+\Delta R)\cos \alpha =r-R\cos \gamma  \label{57}
\end{equation}

Consequently Eq. (55) becomes:

\begin{equation}
\dot r=(H+\Delta H)(r-R\cos \gamma )+HR\cos \gamma  \label{58}
\end{equation}
which immediately gives the same solution

\begin{equation}
\mathbf{\dot r=Hr+\Delta H\cdot (r-R}\cos \mathbf{\gamma )}  \label{59}
\end{equation}

\section{INTERPRETATION OF THE TOTAL $\Delta H$}

In order to clarify the meaning of the total $\Delta H$ of the finite
difference equation (59), we must now return to the obtained results of the
previous sections. Then, we must remember that $R=R_{MW}$ is the $MW$
distance from VC and $R+\Delta R$ the distance of a generic
galaxy/group/cluster from VC, both referring to the epoch of the light
emission, and consequently $\Delta R$ ($\cong -r\cos \gamma $ in the
nearby), still in light-space, represents the radial space separation of
that past epoch ; seemingly $\Delta H$ should be $H(R+\Delta R)-H(R)=\Delta
H_{ga}(r)$ , corresponding to the differential $\Delta R$, and $H(R)$ should
be our Galaxy Hubble constant $H_{MW}$ , both at the past epoch measured by
the light-space $r$. Of course, in this context the $\dot r$ value of Eq.
(59) refers to the instant of the emission epoch. But the available observed 
$\dot r$ is different because it is registered by us at our epoch, and so it
holds, owing to the light delay, the effects of the radial expansion
variation occurred in the time elapsed during the light travel, that is
after the light space $r$. Consequently our $\dot r$ in (59), when
considered as observed $\dot r_{obs}$ , needs to be implemented by a total $%
\Delta H$ representing as a whole the difference between the radial
expansion of the observed galaxy/group/cluster at the emission epoch $%
(r=r_{ga})$ and the one of our Galaxy at the present time $\left( r=0\right) 
$ .

In other words the true $\Delta H$ able to generate the observed velocity $%
\dot r_{obs}$ must be a combination of the Milky Way $\Delta H_{MW}$ , which
is tied to a finite difference of time, plus the above cited $\Delta H_{ga}$
, which instead represents a finite difference of the density/rotation
function $\rho _{*}$ at a precise moment of the past (cfr. Eq. (15)).

To show mathematically that above explained we have to calculate the finite
difference $\Delta H$ as follows: 
\begin{equation}
\Delta H=H_{ga}(r)-H_{MW}(0)=H_{ga}(r)-H_{MW}(r)+H_{MW}(r)-H_0=\Delta
H_{ga}(r)+\Delta H_{MW}(r)  \label{60}
\end{equation}

The previous (60) has general validity, but the value of $\Delta H_{ga}$ is
crucial because its presence would mean the Universe being anisotropic.

First, let us try to carry out a differential analysis limited to the nearby
Universe.

Taking into account the Hubble function $H_{ga}(r,R_{MW}(r)+\Delta R(r))$ ,
relative to a galaxy observed as far as $r$ from us and $R_{MW}(r)+\Delta
R(r)$ from VC, $R_{MW}(r)$ being the Milky Way distance from VC at the epoch
of the light emission, and the $H_{MW}(r,R_{MW}(r))$, relative to the Milky
Way always at the epoch $r$, and the contemporary $H_0=H_0(0,R_0)$ of our
Milky Way, we can apply twice the Taylor series, in succession as follows: 
\begin{equation}
H_{ga}(r,R_{MW}(r)+\Delta R(r))=H_{MW}(r,R_{MW}(r))+\left( \frac{\partial
H_{ga}}{\partial R}\right) _{R=R_{MW}(r)}\cdot \Delta R(r)+...  \label{61}
\end{equation}
\begin{equation}
H_{MW}(r,R_{MW}(r))=H_{MW}(r)=H_0+\left( \frac{\delta H_{MW}(r)}{\delta r}%
\right) _{r=0}\cdot r+...  \label{62}
\end{equation}
from which it results the final one with two total derivatives, that is 
\begin{equation}
H_{ga}(r,R_{MW}(r)+\Delta R(r))=H_0+\left( \frac{\delta H_{ga}}{\delta r}%
\right) _{r=0}\cdot r+\left( \frac{\partial H_{ga}}{\partial R}\right)
_{R=R_{MW}(r)}\cdot \Delta R(r)+...  \label{63}
\end{equation}

Then our searched $\Delta H=H_{ga}(r,R_{MW}(r)+\Delta R(r))-H_0$ , after the
introduction in (63) of the corresponding derivatives of the Hubble constant
formulation (15), becomes

\begin{equation}
\Delta H=\Delta H_{MW}+\Delta H_{ga}\cong K_0r+Q\Delta R(r)  \label{64}
\end{equation}
where 
\[
K_0=3.17\times 10^{33}\left[ \frac 2{t_0^2}-\frac{H_{0_{s^{-1}}}}{t_0}+\frac
23\pi Gt_0\left( \frac{d\rho _{*_{MW}}}{dt}\right) _{t_0}\right] %
\hspace{.1in}Q=-9.51\times 10^{43}\left[ \frac 23\pi Gt\left( \frac{\partial
\rho _{*_{ga}}}{\partial R_{(cm)}}\right) _{R_{MW}(r)}\right] 
\]

So the $\Delta H$ of (64), referring to our local nearby Universe in the
usual Hubble units, should not be zero; it should have a value depending on
two components, the former of which, $\Delta H_{MW}\cong K_0r$, through the
well defined $K_0$ coefficient in $Km$ $s^{-1}Mpc^{-2},$ represents a
systematic time effect (TE, hereafter), while the latter, $\Delta
H_{ga}\cong Q\Delta R(r)\propto -\Delta \rho _{*_{ga}}$ , in $Km$ $%
s^{-1}Mpc^{-1}$, represents a space effect (SE, hereafter) depending both on
the space position $\Delta R(r)$ of the observed galaxy $(ga)$ at the epoch $%
t_s$ ($=t_0-\frac{r_{cm}}c$) of its light emission and on the density%
\TEXTsymbol{\backslash}rotation function variation in that epoch inside the
hemispheres, having $\Delta R>0$ or $\Delta R<0$ respectively.

To conclude it is important to remark how $Q\propto -\frac{\partial \rho
_{*_{ga}}}{\partial R}$ , if present, may reasonably hold the same algebraic
sign in the nearby environment. In fact, according to the model hypothesis
of a spherical symmetry distribution around the void center VC, it is likely
to have the sign change of $\partial \rho _{*_{ga}}(R)$ together with the
sign change of $\partial R$ when we change hemisphere. In other words the
constant sign of $Q$ should mean that our Galaxy does not find itself in a
peculiar position like that of a density\TEXTsymbol{\backslash}rotation peak
in space. Let us remark how the quantities included in the square
parentheses of the above (64) coefficients are all in $c.g.s.$ units

\section{HUBBLE FLOW IN THE NEARBY UNIVERSE}

This section has the task to focus the equations of the previous
experimental model, in order to make possible their easy interpretation in
the most important samples of available data in literature. The Hubble
ratios to use for the check must be corrected by the motion of the Sun in
the Local Group, in practice due to galactic rotation, with the standard
vector of $300$ $Km/s$ towards $l=90^0,$ $b=0^0$ (cf. Sandage \& Tammann,
1975a). This means we consider Hubble ratios as seen from our Local Group,
or from our Galaxy, the Milky Way, it being almost motionless within its
Group. No correction, however, has to be applied for the motion of the Local
Group in the cosmic microwave background (CMB)(see sub-section 9.2).

\subsection{Experimental formulation}

Indeed, eq. (53) represents the fundamental equation of all the present
research. In the light of $\Delta H=H_{ga}-H_0\cong K_0r+Q\Delta R$ and of
the only linear relation $H=H_{MW}\cong H_0+K_0r$, eq. (53), including the
contribution due to $\dot w$, can be easily processed as Hubble ratio as
follows 
\begin{equation}
\frac{\dot r_{obs}}r\cong \mathbf{H}_0\mathbf{+K}_0\left( 2r-R\cdot \cos
\gamma \right) \mathbf{+Q\Delta R}\left( 1-\frac Rr\cos \gamma \right) +R%
\frac{\dot w}r\sin \gamma  \label{65}
\end{equation}

If now eq. (65) is referred to the very nearby Universe, a strong
perturbative effect by the ratio $\dot w/r$ must be expected, being here the
higher $\dot w$ the smaller $r$. The same effect should reasonably happen to
the coefficient $Q=\left( \frac{\partial H_{ga}}{\partial R}\right)
_{R=R_{MW}(r)}$, whose algebraic sign should be constant according to what
is explained above, below eq. (64). In other words both $Q$ and $\dot w$ in
the nearby environment seem able to produce much of the observed noise.

In order to better clarify the meaning of $Q$ , it is useful to write a
further approximated expression of (65), when $\Delta R\ll R$, that is 
\begin{equation}
\Delta R\cong -r\cos \gamma  \label{66}
\end{equation}

which, ignoring the noise of $\dot w$, transforms (65) in the following 
\begin{equation}
\frac{\dot r_{obs}}{r_{*}}\cong H_0+2K_0r_{*}-(K_0R+Q\cdot r_{*})\cos \gamma
+QR\cos ^2\gamma +...  \label{67}
\end{equation}

The above (67) supplies a roughly quadratic equation in $x=-\cos \gamma $ to
the Hubble ratio $y=\frac{\dot r_{obs}}{r_{*}}$ of a nearby sample of
galaxies all at distance $r_{*}$, through the combination of a dipole and
quadrupole type anisotropy.

\subsection{ Homogeneity \& isotropy impose no quadrupole amplitude}

Let us observe that the quadrupole amplitude $QR$ might be considered in
terms of a true perturbative space density\TEXTsymbol{\backslash}rotation
effect (SE), whose meaning may be connected to variations of the matter
density or of the space rotation of the cosmic sphere centred on the
expansion center, when we consider objects having $\Delta R$ of our
environment. In other words the eventual detection of this SE of the nearby
Universe should allow us to identify the local density or rotation variation
in the sphere.

It now becomes important to remark how the persistence of the matter density
constancy in space could represent only homogeneity, but not isotropy, if we
were in presence of a meaningful differential rotation, even if of local
origin. \textbf{In this case all the quadrupole amplitude should be due to
differential rotation}. Otherwise, if we find such perturbation to be rather
undefined, one should assume the constancy of the density\TEXTsymbol{%
\backslash}rotation function with respect to space, according to the total
homogeneity-isotropy condition which imposes the Universe to have no
quadrupole component; this means assuming zero the value of $\Delta \rho
_{*_{ga}}$, that is of $Q$ in eq. (64). However, even assuming rigorously
true the exclusive homogeneity-isotropy condition, we must in any case
consider the other important physical effect, the one entirely due to the
light delay. This time effect (TE) is systematic and able to generate a true
expansion dipole, having its amplitude measured by the value of $K_0R$ in
the formula 
\begin{equation}
\frac{\dot r_{obs}}r\cong H_{*}-K_0R\cos \gamma  \label{68}
\end{equation}

\subsection{Physical meaning of $H_{*}$ and $K_0R\cos \gamma $}

Eqs. (65)(66)(67) are indeed very important, as their application to the
very nearby Universe permit accurate verification of the model. Furthermore
they show now another important feature, that is the new Hubble parameter $%
H_{*}$, whose physical meaning immediately comes to light. In fact the $%
H_{*} $ of eq. (68), being $H_{*}=H_0+2K_0r=H+K_0r$ and $\dot r_{obs}$ $r$
observed quantities as radial velocity and light-space, is the Hubble ratio $%
\frac{\dot r_{obs}}r$ of observed sources located at $\gamma \cong 90^0$ ,
in terms of the Hubble constant $H$ ($=H_0+K_0r$) at the light emission
epoch, plus the same increment $K_0r$ due to the observer deceleration and
then to the slowing of the expansion, occurring during the time taken for
the light to travel from the source, that produces a relative opposite
effect of observed velocity increasing. In other words the measures from the
earth on our Galaxy are affected both by seeing past epochs and by being
referred to an observer having a decreased expansion velocity with respect
to the time of the light emission. Such an effect is easier to understand
and greatly amplified when we observe sources located at $\gamma \cong
0^0,180^0$ directly along the radial expansion direction, becoming here $%
K_0r-K_0R\cos \gamma $ the resulting drop $(\gamma \cong 0^0)$ or rise $%
(\gamma \cong 180^0)$ of the observed outer galaxies' Hubble ratio.

\subsection{Differential rotation}

The problem of rotation for the Universe is an open question, whose
difficult analysis is particularly related to the physical interpretation
given to it. Indeed a rigid rotation should not be detectable, while a
differential one, as that due to angular momentum conserved or to local
perturbative effects, and belonging to our nearby environment, should
produce visible effects. Indeed the differential rotation, if present, may
be described in the eqs. (53)(65) by to the additional term $R\dot w\sin
\gamma $. Now, if there is differential rotation, one should have opposite
algebraic values of $\dot w$ corresponding to equal $\gamma $ in the same
hemisphere. And if this is the case, the global effect would be that of
scattering of Hubble ratios data referring to the same angles $\gamma $.
Hence a simple way to remove such a systematic noise, as well as other
random effects, may be the use of normal points when there is sufficient
number of data in the numerical analysis. So the new basic assumption, 
\begin{equation}
\langle R\frac{\dot w}r\rangle =0  \label{69}
\end{equation}
adopted for a normal point corresponding to the same value of $\gamma $ and
to the average distance $r_{*}$, is able to produce a normal Hubble ratio
formulation of eq. (65) which might permit the cancellation of the eventual
differential rotation of our nearby Universe.

\section{HUBBLE FLOW ACCORDING TO $K_0=3H_0^2/c$}

Lastly, according to the simulation solution of section 4, let us try to
write the more general finite difference expansion equation, referred to a
more distant Universe .

In this case we can substitute $H$, $\Delta H,$ $R$ in (53) the formulas
(38)(40) derived by the Galaxy Hubble law, plus the finite difference $%
\Delta H_{ga}$ of the (60) expression.

It results:

\begin{eqnarray}
\frac{\dot r_{obs}}r &=&H_0+\frac{3H_0^2}{c-3H_0r}\left[ 2r-R_0\cos \gamma
\left( 1-\frac{3H_0r}c\right) ^{\frac 13}\right] +  \label{70} \\
&&\text{ }+\Delta H_{ga}\left[ 1-\frac{R_0\cos \gamma }r\left( 1-\frac{3H_0r}%
c\right) ^{\frac 13}\right] +\frac{R_0\dot w\sin \gamma }r\left( 1-\frac{%
3H_0r}c\right) ^{\frac 13}  \nonumber
\end{eqnarray}

\section{CONCLUSIONS : A fundamental test}

The conclusions and consequences, that may be drawn from the previous
contents of theory and modelling of the expanding Universe from the huge
void center, are indeed many and full of cosmological and astrophysical
implications. Anyway, prior to enter into any details, it is fundamental to
show experimentally the correctness of the model. On the ground of the
successful check work carried out on a lot of available data samples, what
has to be affirmed, at this advanced stage of the research, is the
conclusive experimental confirmation of the expansion center presence
predicted by the model and of the good formal accuracy shown by the obtained
fundamental equations. In particular both the linear (dipole) and the
quadratic structure (dipole +quadrupole) of the Hubble ratio trend in a
Galaxy entourage of about constant $r$ have had full experimental
confirmation. However the negative constant value which results in the
nearby Universe for the quadrupole amplitude $QR$, or $Q$, does not persist
in the larger-scale environment of the nearby Aaronson (1986) clusters.
Indeed it is even possible a sign change of $Q$. This fact, together with
the verified coincidence of the local expansion solution at different
distances with $Q=0$ assumed (see paper II), at present seems to support a
local origin of the quadrupole amplitude $Q$ of the very nearby Universe.

\subsection{Expansion center check}

Of course, the first fundamental check must give the uniqueness of the
expansion center position.

About the coordinates of the Bahcall \& Soneira void center as expansion
center, both the linear and the quadratic formulation in $\cos \gamma $ ,
applied to our very nearby environment, seem to be able to allow their
confirmation. From a series of (68)(67) least square fittings of data by the
Aaronson et al. $308$ nearby galaxy catalog (1982), processed according to
the contents of their following 1986 paper, at the present time and limited
to this Aaronson data set, a minimum standard deviation value has resulted,
corresponding to a lightly shifted void center position, at $\alpha
_{VC}\approx 9^h.8$ , $\delta _{VC}\approx +18^0$ (dipole solution) and at $%
\alpha _{VC}\approx 9^h.5$ , $\delta _{VC}\approx +20^0$ (dipole+quadruple
solution), that is at about $+0.8^h$ and $+0^h.5$ in right ascension and at
about $-12^0$ and $-10^0$ in declination , respectively, with respect to the
Bahcall \& Soneira huge void center coordinates.

The importance of such result is especially due to the fact that the
resulted expansion center is coinciding with a physical point of the sky. In
fact Bahcall \& Soneira detected the void center location by a different
approach tied to a rich cluster distribution observed in the sky, entirely
independent of the expansion model. On the other hand the above obtained $%
\alpha $ $\delta $ rectification could derive from the distortion produced
by different effects like those listed below. Consequently a refinement of
such coordinates is surely possible, but only through the convergence of
different methods and contributors.

\subsection{Dipole solution: $K_0=3H_0^2/c$ confirmed}

The second fundamental check comes from the application of the dipole
equation (68) to two separate important samples of individual galaxies.
These are the above cited AA1 catalog of 308 individual nearby galaxies and
the other Aaronson et al. (1986) sample (AA2) of 148 more distant individual
galaxies. Both the samples may be considered homogeneous and rich enough ,
even if affected by large scattering in distance. Indeed, being the average
individual distance of AA2 about 5 times greater than that of AA1, a
solution can be attempted. To clarify the procedure, we can check two
diagrams (see Fig. 5 and Fig. 6 of the optional section in Paper II: Mini
check atlas of Hubble ratio dipoles), where the corresponding observed
Hubble ratios $\dot r_{obs}/r$ of both the samples are plotted against the
function $-\cos \gamma $ computed with respect to the original Bahcall \&
Soneira void center coordinates. The least square method applied to the
algebraic system of the 308 and 148 dipole equations (68), respectively,
gives the solutions listed in the small table below, where $s$ is the
standard deviation of the fit and all the quantities are in Hubble units.

\begin{center}
\begin{tabular}{|l|l|l|l|l|}
\hline
Sample & $\langle r\rangle _{sample}$ & $H_{*}=H_0+2K_0\langle r\rangle $ & $%
K_0R$ & $s$ \\ \hline
308AA1 & $16.13$ & $90.61\pm 1.80$ & $16.37\pm 2.99$ & $28.67$ \\ \hline
148AA2 & $70.69$ & $98.79\pm 1.75$ & $15.42\pm 2.91$ & $19.79$ \\ \hline
\end{tabular}
\end{center}

The previous table generates a simple algebraic system to solve, whose
solution, with errors let aside, allows to find the following:

\begin{equation}
H_0=88.2\hspace{.3in}K_0=0.075\hspace{.3in}R_{AA1}=218\hspace{.3in}%
R_{AA2}=206  \label{71}
\end{equation}

This solution gives an accurate confirmation of the correlation between $K_0$
and $H_0$ , as foreseen by the simulation result of Eq. (35) in section 4 $%
(3\times 88.2^2/299800=0.078)$. At the same time such solution is clearly
affected by the Hubble ratio scattering, which is too high. Indeed the noise
is necessarily great owing to a lot of combined effects, whose nature has
been partially described in the previous sections, as connected to the
presence of $\dot w$. Now, without entering here into details, we can
qualitatively conclude that the consistent registered deviations are due to
at least 8 different effects. i.e. :

\begin{center}
1) large scattering in distance inside the samples of galaxies;

2) random errors in the Hubble ratio measurements;

3) systematic errors in the Hubble ratios;

4) rough expansion center coordinates;

5) presence of a quadrupole perturbation;

6) perturbative actions inside groups and clusters;

7) possible limits in using the first order coefficient: $K_0$ ;

8) possible presence of differential rotation around the expansion center.

\hspace{1.0in}
\end{center}

A few of these effects may be removed or reduced, applying a more precise
equation than Eq. (68) and, especially, passing to combined Hubble ratios of
groups and clusters. In this case the effects (1)(2)(6) tend to vanish, like
(7) in the very nearby environment, while on the other hand (5) and (8)
should persist and become easier to study. Anyway it is important to remark
that a large part of the noise is probably due to a mixing of precise
systematic physical effects, and not to a generic chaotic thermic
distribution.

In conclusion a few words have to be spent on an unconsidered effect among
the 8 listed. We refer to the motion of the Local Group (LG) in the cosmic
microwave background (CMB). In fact it must be said that the AA1 dipole here
presented , after adding the correction of such a motion, that is by means
of the entire application of $M(Sun)=B(Sun)+C(LG)$ $(=300$ $Km/s$ to $%
l=90^0, $ $b=0^0$ $+$ $610$ $Km/s$ to $l\cong 268^0,$ $b\cong +27^0)$ to the
Sun's velocity in the CMB, practically vanishes together with the
correlation between $K_0$ and $H_0$ , generating a standard deviation
amplified more than double ($s=75.61$ $!)$. Such exploratory check seems to
find a reasonable explanation only by assuming that the motion of the Galaxy
or Local Group in the cosmic microwave background might belong also to the
nearby galaxies\TEXTsymbol{\backslash}groups, in a sort of large flow
running almost along the same direction. If this were the case, the Galilean
relativity effect, while one is observing the nearby Universe, results in
the complete cancellation of the motion, while on the other hand its
consideration, exclusively in terms of applied correction to our Local
Group, enormously increases the noise, at the same time strongly involving
the present solution.

\newpage\ 

\textbf{REFERENCES}

Aaronson, M. et al. 1982, Ap. J. Suppl. Series, 50, 241

Aaronson, M. et al. 1986, Ap. J. 302, 536

Alpher and Herman, 1948

Ambartsumian, V. A. 1961, A. J., 66, 536

Bahcall, N.A., 1988, Ann. Rev. Astron. Astrophys. 26, 631

Bahcall, N.A. and Soneira, R.M. 1982, Ap. J. 262, 419

Faber, S.M. et al. 1989, Ap. J. Suppl. Series 69, 763

Gamow, 1948

Geller, M.J. and Huchra, J.P. 1989, Science 246, 897

Hubble, E. 1929, Proc. Nat. Acad. Sci., 15, 168

Kirshner,R.P.,Oemler, A., Jr.,Schechter, P.L.and Schectman,S.A. 1981,
Ap.J.(Letters), 248,L57

Lahav, O. 1987, Mon. Not. R. astr. Soc. 225, 213

Lipovetsky, V.A. 1987, Proc. Semin. Large Scale Structure of the Universe,
1986,

\hspace{1in}p. 47. Stavropol, USSR: Spec. Astrophys. Obs.

Lema\^itre, G., 1933a, Acad. Sci., Paris, Comptes Rend. 196, 1085

\hspace{.85in}1933b, Acad. Sci., Paris, Comptes Rend. 196, 903

Lorenzi, L. 1989, 1991, Contributi N. 0,1, Centro Studi
Astronomia-Mondov\`\i , Italy

\hspace{.7in}1995, Mem. S.A.It., V. 66, N. 1, p. 249 (1993-Asiago Proc. II
Cosmology)

\hspace{.7in}1995bc, Sesto Pusteria International Workshop Book, eds.-SISSA
ref. 65/95/A

\hspace{.7in}1996, Astro. Lett. \& Comm., 33, 143 (1994-Grado3 Proc.,
eds.-SISSA ref. 155/94/A)

\hspace{.7in}1999-II (parallel paper)

Lynden-Bell, D. et al. 1988, Ap. J. 326, 19

Milne, 1933

Nottale,L., Pecker,J.C., Vigier,J.P., et Yourgrau,W., 1976, La Recherce N.
68, V. 7, 529

Penzias, A.A., and Wilson, R.W. 1965, Ap. J., 142, 419

Rubin, V.C., Ford, W.K., and Rubin, J.S. 1973, Ap. J. Letters 183, L111 (RFR)

Sandage, A. and Tammann, G.A. 1975, Ap. J. 196, 313 (Paper V)

\hspace{1.2in}1975b, Ap. J. 197, 265 (Paper VI)

Smooth, G.F. et al. (COBE) 1992, preprint-submitted to Ap. J. Letters)

\newpage\ 

\textbf{CAPTIONS OF THE FIGURES}

\vspace*{.1in}

\textbf{Figure 1 (Paper I)}

Local cosmographic section, as described in the vectorial verification of
the text.

\vspace{.1in}

\textbf{Figure 2 (Paper II)}

Hubble ratios of 52 groups by de Vaucouleurs (1965) plotted

against the $-\cos \gamma $ of each group.

\vspace{.1in}

\textbf{Figure 3 (Paper II)}

Hubble ratios of 20 groups by Sandage \& Tammann (1975) plotted

against the $-\cos \gamma $ of each group (cf. Table 1).

\vspace{.1in}

\textbf{Figure 4 (Paper II)}

The observed Hubble ratios of 83 individual galaxies by Sandage \& Tammann

(Tables 2-3-4:1975-V) plotted against the function $-\cos \gamma $ of each
corresponding galaxy.

\vspace{.1in}

\textbf{Figure 5 (Paper II)}

The observed Hubble ratios of 308 nearby galaxies by Aaronson et al.

(1982) plotted against $-\cos \gamma _{galaxy}$ of each corresponding galaxy.

\vspace{.1in}

\textbf{Figure 6 (Paper II)}

The observed Hubble ratios of 148 more distant individual galaxies by
Aaronson et al. (1986)

plotted against the function $-\cos \gamma _{cluster}$ of the corresponding
cluster.

\vspace{.1in}

\textbf{Figure 7 (Paper II)}

Hubble ratios of 31 groups by Aaronson et al. (1982) plotted

against the $-\cos \gamma $ of each group (cf. Table 2).

\vspace{.1in}

\textbf{Figure 8 (Paper II)}

Hubble ratios of 10 clusters by Aaronson et al. (1986) plotted

against the $-\cos \gamma $ of each cluster (cf. Table 3).

\vspace{.1in}

\textbf{Figures 2,3,7,8 (Paper II)}

The sizes of the plotted points are in proportion to the member number, n$%
_{obs}$,

of the group\TEXTsymbol{\backslash}cluster, according to the following:

$\Rightarrow n_{obs}\leq 4;$\hspace{.4in} $\Rightarrow 4<n_{obs}\leq 8;%
\hspace{.4in}\Rightarrow 8<n_{obs}\leq 16;\hspace{.4in}\Rightarrow
16<n_{obs} $

\end{document}